\documentclass[aps,prl,twocolumn,showpacs,superscriptaddress,groupedaddress]{revtex4}
\usepackage{color}
\usepackage{xcolor}
\usepackage[]{graphicx}
\usepackage{latexsym}
\usepackage{natbib}
\usepackage{amssymb}
\usepackage{amsmath}
\usepackage[caption=false]{subfig}
\usepackage{multirow}

\usepackage{dsfont}

\usepackage[final]{feynmp}
\newcommand{\h}{\hspace{1pt}}

\newcommand{\s}{\sigma}

\newcommand{\la}{\langle}
\newcommand{\ra}{\rangle}

\newcommand{\be}{\begin{equation}}
\newcommand{\ee}{\end{equation}}
\newcommand{\bc}{\begin{cases}}
\newcommand{\ec}{\end{cases}}

\begin{document}

\title{Emergence of Robustness in Network of Networks}

\author{Kevin Roth}
\affiliation{Levich Institute and Physics Department, City College of New York, New York, NY 10031}
\affiliation{Theoretical Physics, ETH Z\"urich, 8093 Z\"urich, Switzerland}
\author{Flaviano Morone}
\affiliation{Levich Institute and Physics Department, City College of New York, New York, NY 10031}
\author{Byungjoon Min}
\affiliation{Levich Institute and Physics Department, City College of New York, New York, NY 10031}
\author{Hern\'an A. Makse}
\email[]{hmakse@lev.ccny.cuny.edu}
\affiliation{Levich Institute and Physics Department, City College of New York, New York, NY 10031}


\begin{abstract}
A model of interdependent networks of networks (NoN) has been
introduced recently in the context of brain activation to identify the neural collective influencers in the brain NoN.
Here we develop a new approach to derive an exact
expression for the random percolation transition in 
Erd\"{o}s-R\'enyi NoN. Analytical calculations are in excellent
agreement with numerical simulations and highlight the robustness of
the NoN against random node failures. Interestingly, the phase
diagram of the model unveils particular patterns of interconnectivity
for which the NoN is most vulnerable. Our results help to understand the emergence of robustness in such interdependent architectures.
\end{abstract}

\maketitle

%------------------------------------------------------------------------%
%                        INTRODUCTION                                   %
%------------------------------------------------------------------------%
\hspace{0pt} %WORKAROUND TO ALIGN BOTH COLUMNS!
\vspace{-2\baselineskip}

Many biological, social and technological systems are composed of
multiple, if not vast numbers of, interacting elements. In a stylized
representation 
each element is portrayed as a node
and the interactions among nodes as mutual links, so as to form what is 
known as a network~\cite{Newman2010:Book}. A finer description
further isolates several sub-networks, called modules,
each of them performing a different function. These modules are, in turn, 
integrated to form a larger aggregate referred to as a network of
networks (NoN).
A compelling problem is how to define the interdependencies between modules, specifically how the functioning of nodes in one module
depends on the functioning of nodes in other
modules~\cite{Buldyrev2010:Cascade, Gao2012:Interdependent,Bianconi2015:MutualComponent, ReducingCouplingStrength,saulo}.

Current models of such interdependent NoN, inspired by the power grid,
represent dependencies across  modules through very fragile
couplings~\cite{Buldyrev2010:Cascade, Gao2012:Interdependent}, 
such that the random failure of few nodes gives rise to a
catastrophic cascading collapse of the 
NoN. Many real-life systems, however, exhibit high resilience against
malfunctioning. The
prototypical example of such robust modular architectures is the
brain, which thus cannot fit in catastrophic NoN
models~\cite{saulo}.\,\ To cope with the fragility of current NoN
models, we recently introduced a model of interdependencies in
NoN~\cite{preprint}, inspired by the phenomenon of top-down
control in brain activation~\cite{gallos,sigman}, in order to study the
impact of rare events, i.e.\ non-random {\it optimal percolation}
\cite{MoMa}, on the global communication of the brain with
application to neurological disorders.  

Here we investigate the robustness of this NoN model with respect to
typical node failures, i.e.\ {\it random percolation}. More precisely,
we develop a new approach to derive an analytical expression for the
random percolation phase diagram in Erd\"{o}s-R\'enyi
(ER) NoN, which predicts the conditions responsible for the emergence of robustness and the absence
of cascading effects.

%------------------------------------------------------------------------%
%                                DEFINITION                                  %
%------------------------------------------------------------------------%
%-----------------------------------------------%
%                     Figure  1                      %
%-----------------------------------------------%
\begin{figure}%[t]
\includegraphics[width=0.95\columnwidth]{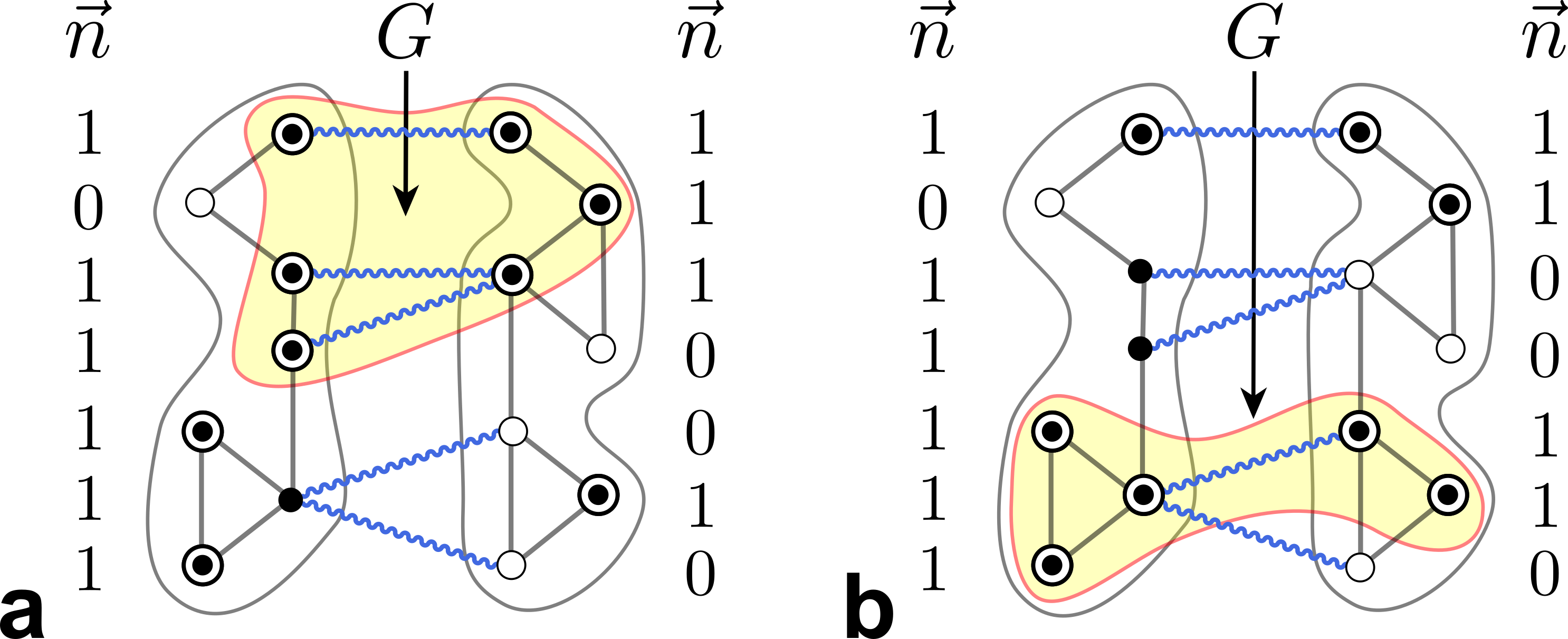}
\caption{(Color online) \textbf{Robust interdependent 2-NoN.}
  Intra-module links (black) represent connectivity, while
  inter-module links (wiggly blue lines) express mutual dependencies.  The
  occupation variable $n_i $ specifies whether a node is present 
	($n_i = 1$) or removed ($n_i = 0$).  The activation state $\s_i$, defined
  through inter-module dependencies, indicates whether a node is
  activated ($\s_i = 1$) or inactivated ($\s_i = 0$). 
	Nodes can be activated even if they do not belong to the giant
  connected activated component $G$. Note also that the configuration of
  occupation variables $\vec{n}$ is identical for the module on the
  left in \textbf{a} and \textbf{b}. Legend:
 % $\vcenter{\hbox{
	\protect\includegraphics[width=0.7em]{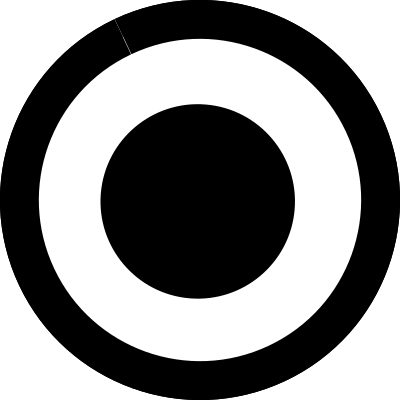} $\s_i=1$;
  $\bullet$ $n_i=1$, $\s_i=0$; $\circ$ $n_i=0$, $\s_i=0$.}
\label{fig:fig1}
\end{figure}

{\bf Definition of control intra-modular links.---} Consider $N$ nodes
in a NoN composed of several interdependent modules\ 
{(Fig.\,\ref{fig:fig1})}.\ We distinguish the roles of intra-module
links connecting nodes within a module, and inter-module dependency
links (corresponding to control links in the brain \cite{gallos,
  saulo}), connecting nodes across modules: the former (intra-links)
only represent whether or not two nodes are connected, the latter
(inter-links) express mutual control.  Every \mbox{node $i$}
has $k^{\rm in}_{i}$ intra-module links, referred to as node $i$'s
in-degree, and $k^{\rm out}_{i}$ inter-module connections, referred to
as $i$'s out-degree.

Each node can be present or removed, and, if present, it can be
activated or inactivated. We introduce the binary occupation
variable\ $n_i = 1, 0$\ to specify whether node~$i$ is present $(n_i =
1)$ or removed $(n_i = 0)$. By virtue of inter-module dependencies,
the functioning of a node in one module depends on the functioning of
nodes in other modules.  In order to conceptualize this form of
control, we introduce the activation state $\s_i$, taking
values $\s_i = 1$ if node $i$ is activated and $\s_i = 0$ if not. A
node~$i$ with one or more inter-module dependency/control connections
$(k^{\rm out}_{i} \geq 1)$ is activated $(\s_i = 1)$ if and only if it
is present $(n_i = 1)$ and at least one of its out-neighbors~$j$ is
also present $(n_j = 1)$, otherwise it is not activated $(\s_i =
0)$. In other words, a node with one or several inter-module
dependencies is inactivated when the last of its out-neighbors is
removed.

The rationale for this control rule is that the activation
($\s_i=\s_j=1$) of two nodes connected by, for instance, one
inter-link occurs only when both nodes are occupied, $n_i=n_j=1$.
If just one of them is unoccupied, let's say $n_j=0$, then
both nodes become inactive. Thus, $\s_i=0$ even though $n_i=1$, and we
say that $j$ exerts a control over $i$.  This rule models the way
neurons control the activation of other neurons in distant brain
modules via control/dependency links (fibers through the white matter)
in a process known as top-down influence in sensory processing
\cite{sigman}.  Mathematically, $\s_i$ is defined as
\begin{equation}
\s_i\ =\ n_i\bigg[1-\prod_{j \in \mathcal{F}(i)}(1-n_j)\bigg]\,,
\label{eq:sigma}
\end{equation} 
where $\mathcal{F}(i)$ denotes the set of nodes connected to $i$ via
an inter-module link.  
Conceptually, the inter-links define a
mapping from the configuration of occupation variables $\vec{n} \equiv
(n_1,...,n_N)$ to the configuration of activated states $\vec{\s}
\equiv(\s_1,...,\s_N)$, as given by Eq.\,(\ref{eq:sigma}).

Not all nodes participate in the control of other nodes via
dependencies, i.e.\ a certain fraction of them does not establish
inter-links. 
If a node does not have inter-module dependencies, it activates as long  
as it is present:
\begin{equation}
\s_{i} = n_{i}\,,\,\,\,\,\,\, \mbox{for $k^{\rm out}_{i}=0$ .}
\label{eq:sigma0}
\end{equation}
Therefore, products over empty sets $\mathcal{F}({i})=\emptyset$
default to zero in Eq.\,(\ref{eq:sigma}).  This
last property also guarantees that we recover the single network
case for vanishing inter-module connections $(\langle k^{\rm
  out}_{i}\rangle \rightarrow 0$), i.e.\ when considering the limiting
case of one isolated module only.

When a fraction of nodes is removed, the NoN breaks into isolated
components of activated nodes. In this work we focus on the
\textit{largest (giant) mutually connected activated component} $G$,
which encodes global properties of the system.
In contrast to previous NoN models \cite{Buldyrev2010:Cascade,
  Gao2012:Interdependent}, in our model a node can be activated even
if it does not belong to $G$ (see Fig.\,\ref{fig:fig1}). 
Indeed, the activation of a node, given by Eq.\,(\ref{eq:sigma}), is not tied to its
membership in the giant component.
{\color{black}Therefore, a node can be part of $G$ without being part of the largest connected activated component in its own module (consider for instance the top left node in Fig.\,1\,{\bf a}).}\ 
As a consequence, controlling dependencies in the NoN do not lead
to cascades of failures, which ultimately explains the robustness of
our NoN model. 
{\color{black}
In the model of Refs.\,\cite{Buldyrev2010:Cascade,
  Gao2012:Interdependent}, on the other hand, a node can be activated (therein termed ``functional'') if and only if it belongs to the largest connected component of its own module and (for the case that it has inter-module dependency links) its out-neighbors also belong to the giant component within their module.} 
Indeed, in Refs.\,\cite{Buldyrev2010:Cascade,Gao2012:Interdependent} the propagation
of failures is not local as in Eq.\,(\ref{eq:sigma}), implying that the failure of a
single node may catastrophically destroy the NoN.

In order to quantify robustness, we measure the impact of node
failures $n_i=0$ on the size of $G$
\cite{Buldyrev2010:Cascade,Gao2012:Interdependent,Bianconi2015:MutualComponent}.
More precisely, we calculate $G$ under typical configurations
$\vec{n}$, sampled from a flat distribution with a given fraction $q
\equiv 1 - \sum_{i = 1}^{N}n_i/N$ of removed nodes, and show that $G$
remains sizeable even for high values of $q$. In practice, starting
from $q = 0$, we compute $G(q)$ while progressively increasing the
fraction $q$ of randomly removed nodes. 
The robustness of the NoN is
then formally characterized by the critical fraction $q_c$, the
percolation threshold, at which the giant connected activated
component collapses $G(q_c) = 0$ \cite{Buldyrev2010:Cascade,
  Gao2012:Interdependent}. Accordingly, NoN models with high $q_c$
(ideally close to 1) are robust, whereas low $q_c$ is considered
fragile.
{\color{black}A plot of $G(q)$ for ER 2-NoN is shown in the inset of Fig.\,2.}

%------------------------------------------------------------------------%
%                          MESSAGE PASSING                          %
%------------------------------------------------------------------------%
{\bf Message Passing.---} The problem of calculating $G$ can be solved
using a message passing approach
\cite{Bianconi2015:MutualComponent,MoMa,Zdeborova2014:Percolation} 
which provides exact solutions on locally tree-like NoN, containing a small number of short loops \cite{Zdeborova2014:Percolation}. 
This includes the thermodynamic limit $(N\to\infty)$ of Erd\"{o}s-R\'enyi and scale-free random graphs as well as the configuration model (the maximally random graphs generated from a given degree distribution), which contain loops whose typical length grows logarithmically with the system size \cite{Dorogovtsev2003}. 

In principle, it works like this: each node receives messages from its
neighbors containing information about their membership in $G$.  Based
on what they receive, the nodes then send further messages until
everyone eventually agrees on who belongs to $G$.  In practice, we
need to derive a self-consistent system of equations that specifies
for each node how the message to be sent is computed from the incoming
messages {\cite{Mezard2009}}.
To this end, we introduce two types of messages:
$\rho_{i\to j}$ running along an intra-module link and $\varphi_{i\to j}$ running along an inter-module link.
Formally, we denote $\rho_{i\to j}\equiv$ \textit{probability that node $i$ is connected to $G$ other than via in-neighbor $j$}, 
and $\varphi_{i\to j}\equiv$ \textit{probability that node $i$ is connected to $G$ other than via out-neighbor $j$}. 
The binary nature of the occupation variables and the
activation states constrains the messages to take values
$\rho_{i\to j}, \varphi_{i\to j} \in \{0,1\}$.

A node can only send non-zero information if it is activated,
hence the messages must be proportional to~$\s_{i}$. 
Assuming node $i$ is activated, it can send a
non-zero intra-module message $\rho_{i\to j}$
to node $j$ if and only if it receives a non-zero message by at least
one of its in-neighbors other than $j$ \textit{or} one of its
out-neighbors.  Similarly, we can consider the message $\varphi_{i\to
  j}$ along an inter-module link.  Thus, the self-consistent system of
message passing equations is given by:
\begin{align}
\rho_{i\to j} &= \s_{i} \Big[ 1 - \hspace{-.2cm}\prod_{k \in \mathcal{S}(i) \setminus j} \hspace{-.2cm} (1-\rho_{k\to i})\hspace{-.1cm} \prod_{k \in \mathcal{F}(i)} \hspace{-.1cm}( 1 - \varphi_{k\to i} ) \Big]\ ,\label{eq:messagePassingRho} \\
\varphi_{i\to j} &= \s_{i} \Big[ 1 - \hspace{-.1cm}\prod_{k \in \mathcal{S}(i)} \hspace{-.1cm} (1-\rho_{k\to i} )\hspace{-.2cm} \prod_{k \in \mathcal{F}(i)\setminus j} \hspace{-.2cm}( 1 - \varphi_{k\to i} ) \Big]\ ,\label{eq:messagePassingVarphi}
\end{align}
where $\mathcal{S}(i)$ denotes the set of node $i$'s intra-module nearest 
neighbors and $\mathcal{F}(i)$ denotes the set of $i$'s inter-module 
nearest neighbors. Note that products over empty sets
$\mathcal{S}(i)=\emptyset$ or $\mathcal{F}(i)=\emptyset$ default to
one. 

In practice, the message passing equations are solved iteratively.
Starting from a random initial configuration $\rho_{i\to j}, 
\varphi_{i\to j} \in \{0,1\}$, the messages are updated until they finally converge.  From the
converged solutions for the messages we can then compute the marginal
probability $\rho_{i} = 0,1$ for each node $i$ to belong to the giant
connected activated component $G$:
\begin{equation}
\rho_{i}\ =\ \s_{i} \Big[\h 1\ - \hspace{-.1cm}\prod_{k \in \mathcal{S}(i)} \hspace{-.1cm} (1-\rho_{k\to i} ) \hspace{-.1cm} \prod_{k \in \mathcal{F}(i)} \hspace{-.1cm}( 1 - \varphi_{k\to i} )\h \Big]\ .
\label{eq:marginal}
\end{equation}

The size of $G$, or rather the fraction of nodes belonging to $G$, can then simply be computed by summing the probability marginals $\rho_i$ and dividing by the system size: $G(\vec{n}) =  \big(\sum_{i = 1}^{N} \rho_{i}\big)/N$.

%------------------------------------------------------------------------%
%                PERCOLATION THRESHOLD                      %
%------------------------------------------------------------------------%
{\bf Percolation Phase Diagram for ER NoN.---}
In what follows we derive an exact expression for the
percolation threshold in Erd\"{o}s-R\'enyi \mbox{2-NoN}, defined as
two randomly interconnected ER modules. Each module is an ER random
graph with Poisson degree distribution, 
$\mathds{P}_z[ k^{\rm in}] = e^{-z}z^{k^{\rm in}}/k^{\rm in}!$ 
for $k^{\rm in}\in\mathbb{N}_0$, 
where $z \equiv \la k^{\rm in}\ra $ denotes the average in-degree.
{\color{black}Similarly, we consider the inter-module links to form a bipartite ER random graph
with Poisson degree distribution},
$\mathds{P}_w[ k^{\rm out}] = e^{-w}w^{k^{\rm out}}/k^{\rm out}!$ 
for $k^{\rm out}\in\mathbb{N}_0$, 
where $w \equiv \la k^{\rm out}\ra $ denotes the average out-degree.
The corresponding distributions for the in-/out-degree at the end of an \mbox{intra-/} inter-link are
given by, 
$\mathds{Q}_z[ k^{\rm in}] = (k^{\rm in}\mathds{P}_z[ k^{\rm in}] \mathds{1}_{\{k^{\rm in}>0\}})/z$ 
and
$\mathds{Q}_w[ k^{\rm out}] = (k^{\rm out}\mathds{P}_w[ k^{\rm out}] \mathds{1}_{\{k^{\rm out}>0\}})/w$,  
for $k^{\rm in}, k^{\rm out}$ in $\mathbb{N}_0$, 
where $\mathds{1}_{\{\cdot\}}$ denotes the indicator function.

{\color{black}The random percolation process is then defined by removing each node
in the NoN independently with probability $q$, which is equivalently
formulated as taking the configurations \mbox{$\vec{n} = (n_1,...,
  n_N)$} at random from the binomial distribution, 
$\mathds{P}_p[\vec{n}] = \prod_{i = 1}^{N} p^{n_i}(1-p)^{1-n_i}$, 
where $p=1-q$ denotes the occupation probability.

The probability of a node to be activated when a randomly chosen fraction $p$ of nodes in the NoN is present, 
$\big\langle \sigma_i \big\rangle_{p} = p\mathds{1}_{\{k_i^{\rm out}=0\}} + p\big[1 - (1 - p)^{k^{\rm out}_i}\big]\mathds{1}_{\{k_i^{\rm out}>0\}}$,
can straightforwardly be obtained by averaging $\sigma_i$, given by Eq.\,(\ref{eq:sigma}), over  $\mathds{P}_p[\vec{n}]$. 
The expected fraction of activated nodes $\big\langle \sigma_i \big\rangle_{p, w} = p\big[1+ e^{-w} - e^{-w p}\big]$
is then given by averaging $\big\langle \sigma_i \big\rangle_{p}$ over $\mathds{P}_w[k^{\rm out}_i]$.
Unlike a node's probability to be present $\langle n_i \rangle_{p} = p$, the probability to be activated $\langle \sigma_i \rangle_p$ is therefore highly dependent on the node's out-degree $k^{\rm out}_i$.
In other words, the deactivations %($\sigma_i=1\rightarrow\sigma_i=0$) 
are highly degree dependent, even if the fraction $q$ of nodes to be removed from the NoN %($n_i = 1 \rightarrow n_i = 0$) 
is chosen randomly!
}

To compute the expectation of messages within the ensemble of ER 2-NoN,
we average the expressions for $\rho_{i\to j}$ and $\varphi_{i\to j}$,
representing the converged solutions to the message passing equations, 
over all possible realizations of randomness inherent in the above distributions. 
In doing
so, we must however make sure to properly account for the fact that,
for \mbox{nodes $i$} with inter-links ($k_i^{\rm out}\geq 1$), the
binary occupation variable $n_i$ shows up more than once within
the entire system of message passing equations, due to the activation
rule for $\s_i$.
Indeed, since the occupation variable is a binary number $n_i \in \{0,
1 \}$, powers of $n_i^{k} = n_i$ for each exponent \mbox{$k \in
  \mathbb{N^{+}}$} and therefore the self-consistency is not affected by the
existence of multiple $n_i$ per node. Yet, when naively averaging with
the distribution of configurations, 
we would incorrectly obtain $n_i^{k}
\overset{\mathds{P}_p}{\longrightarrow} p^{k}$ instead of
$n_i^{k} \overset{\mathds{P}_p}{\longrightarrow} p$, without
properly accounting for the binary nature of the occupation variable
across the entire system of equations.

{\color{black}Specifically, when inserting the expression for the message
$\varphi_{k\to i}$, determined by Eq.\,(\ref{eq:messagePassingVarphi}), into
the expression for $\rho_{i\to j}$, given by Eq.\,(\ref{eq:messagePassingRho}), 
then the activation state $\s_k = n_k [1 - (1-n_i)\prod_{\ell \in
    \mathcal{F}(k)\setminus i}(1-n_\ell) \big]$ (within $\varphi_{k\to
  i}$) reduces to $n_k$, since \mbox{$n_i(1-n_i) = 0$} for binomial variables.
%Without replacing $\s_i$ with $n_i$, as done in Eq.\,(\ref{eq:modifiedMP}), 
%we would erroneously obtain $n_i (1-n_i)
%\overset{\mathds{P}_p}{\longrightarrow} p (1-p)$ when separately
%averaging the expressions for $\rho_{i\to j}$ and $\varphi_{i\to j}$.
In other words, we need to replace $\s_k$ ($\s_i$) with $n_k$ ($n_i$) within the expression for $\varphi_{k\to i}$ ($\varphi_{i\to j}$, Eq.\,(\ref{eq:messagePassingVarphi})).}

Thus, the modified message passing equations we need to average read:
\begin{equation}
\begin{aligned}
\rho_{i\to j} &= \s_{i} \Big[ 1 - \hspace{-.3cm}\prod_{k \in
    \mathcal{S}(i) \setminus j} \hspace{-.2cm} (1-\rho_{k\to
    i})\hspace{-.1cm} \prod_{k \in \mathcal{F}(i)} \hspace{-.1cm}( 1 -
  \varphi_{k\to i} ) \Big]\ ,\\ \varphi_{i\to j} &= n_i \Big[ 1
  - \hspace{-.1cm}\prod_{k \in \mathcal{S}(i)} \hspace{-.1cm}
  (1-\rho_{k\to i} )\hspace{-.2cm} \prod_{k \in
    \mathcal{F}(i)\setminus j} \hspace{-.2cm}( 1 - \varphi_{k\to i} )
  \Big]\ .
\label{eq:modifiedMP}
\end{aligned}
\end{equation}

%We emphasize that the modified message passing equations are not
%self-consistent anymore. Nevertheless, they properly address the
%binary nature of the occupation variable when averaging and therefore
%allow us to recover the correct percolation threshold.

In practice, we expand $\rho_{i \to j}$, given by
Eq.\,(\ref{eq:modifiedMP}), and perform the averaging separately for
each term:
\begin{align}
\rho_{i\to j} &= n_i \Big[1 - \hspace{-.3cm}\prod_{k\in
    \mathcal{S}(i)\setminus j}\hspace{-.2cm}(1-\rho_{k\to i}) \Big]
\mathds{1}_{\{k^{\rm out}_i = 0\}}\\ &+ \s_i \Big[1
  - \hspace{-.3cm}\prod_{k\in \mathcal{S}(i)\setminus
    j}\hspace{-.2cm}(1-\rho_{k\to i})\hspace{-.2cm}\prod_{k\in
    \mathcal{F}(i)}\hspace{-.2cm}(1-\varphi_{k\to i}) \Big]
\mathds{1}_{\{k^{\rm out}_i > 0\}} \ . \nonumber
\label{eq:rho_ij}
\end{align}
The only non-trivial average involves the following expression:
\vspace{.2cm}
\begin{equation}
\begin{aligned}
& \Big\la \s_i \hspace{-.2cm}\prod_{k\in \mathcal{S}(i)\setminus j}\hspace{-.2cm}(1-\rho_{k\to i})\hspace{-.1cm}\prod_{k\in \mathcal{F}(i)}\hspace{-.1cm}(1-\varphi_{k\to i}) \,\mathds{1}_{\{k^{\rm out}_i > 0\}} \Big\ra \\
& = \Big\la n_i \hspace{-.2cm}\prod_{k\in \mathcal{S}(i)\setminus j}\hspace{-.2cm}(1-\rho_{k\to i}) \Big[ \prod_{k\in \mathcal{F}(i)}\hspace{-.1cm}(1-\varphi_{k\to i})\\
& -\ \prod_{k\in \mathcal{F}(i)}\hspace{-.1cm}(1-n_k)(1-\varphi_{k\to i})\Big] \mathds{1}_{\{k^{\rm out}_i > 0\}} \Big\ra \ ,
\end{aligned}
\end{equation}
where we have to account for the fact that
$(1-n_k)(1-\varphi_{k\to i}) = (1-n_k)$.
%, since for the converged messages $n_k \varphi_{k\to i} = \varphi_{k\to i}$.
%(The reason is that if $n_k = 0$ then also $\varphi_{k\to i} = 0$,
%whereas if $n_k = 1$ the statement is trivially true.)
The final expression for the average intra-module message $\rho$
reads:
\begin{equation}
\rho = p \big[ 1+e^{-w}\hspace{-1pt}- e^{-w p}\hspace{-1pt}- e^{-z\rho
    -w}\hspace{-1pt}+ e^{-z\rho -w p}\hspace{-1pt}- e^{-z\rho -w
    \varphi} \big]\hspace{-0.2pt}.
\end{equation}

Averaging the modified inter-link message
$\varphi_{i\to j}$, given by Eq.\,(\ref{eq:modifiedMP}), over all
possible realizations of randomness inherent in the percolation process yields:
\begin{equation}
\varphi\ =\  p\, \big[\, 1 - e^{-z\,\rho\,-w\,\varphi}\, \big]\ .
\end{equation}

The percolation threshold $p_c = 1 - q_c$ of the ER 2-NoN can now be found
by evaluating the leading eigenvalue determining the stability of the
fixed point solution $\{\rho=\varphi=0\}$ to the averaged modified
message passing equations \cite{Zdeborova2014:Percolation}:
\begin{equation}
\left. \left(
\hspace{-2pt}\begin{array}{cc}
\frac{\partial \rho}{\partial \rho} & \frac{\partial \varphi}{\partial \rho}\\
\frac{\partial \rho}{\partial \varphi} & \frac{\partial \varphi}{\partial \varphi}
\end{array}\hspace{-2pt}\right) \hspace{-1pt}\right|_{\{\rho=\varphi=0\}} \hspace{-6pt}= \left(
\hspace{-2pt}\begin{array}{cc}
pz\big[1+e^{-w}\hspace{-2pt}- e^{-wp}\big] & pz\\
pw & pw
\end{array} \hspace{-2pt}\right)_{.}
\end{equation}
The corresponding eigenvalues can readily be~obtained~as 
%$\lambda_{\pm} = \frac{1}{2}\,[\,Tr \pm \sqrt{\,Tr^2 - 4Det\,}\,]$:
\begin{equation}
\lambda_{\pm}\hspace{-2pt}=\hspace{-1pt}\frac{p}{2}\Big[ z[1\hspace{-1pt}+\hspace{-1pt}f]\hspace{-1pt}+w \pm \sqrt{ z^2[1\hspace{-1pt}+\hspace{-1pt}f]^2\hspace{-1pt}+\hspace{-1pt}2zw[1\hspace{-1pt}-\hspace{-1pt}f]\hspace{-1pt}+\hspace{-1pt}w^2} \Big]
\label{eq:leadingEigenvalue}
\end{equation}
where we define $f(p) \equiv e^{-w} - e^{-w p}$.
Formally, the fixed point solution $\{\rho=\varphi=0\}$ is stable if
and only if $\lambda_{+} \leq 1$ 
\cite{MoMa,Zdeborova2014:Percolation}. 
The implicit function theorem then allows us to obtain the percolation 
threshold $p_c=1-q_c$ by saturating the stability condition  
as follows:
\begin{equation}
\lambda_{+}\h(\h p,\h z,\h w\h)\ =\ 1 \hspace{.2cm}\rightarrow\,\,\ p_c\h(\h z,\h w\h)\ .
\label{eq:stabilityCondition}
\end{equation}

\begin{figure}%[t]
\includegraphics[width=\columnwidth]{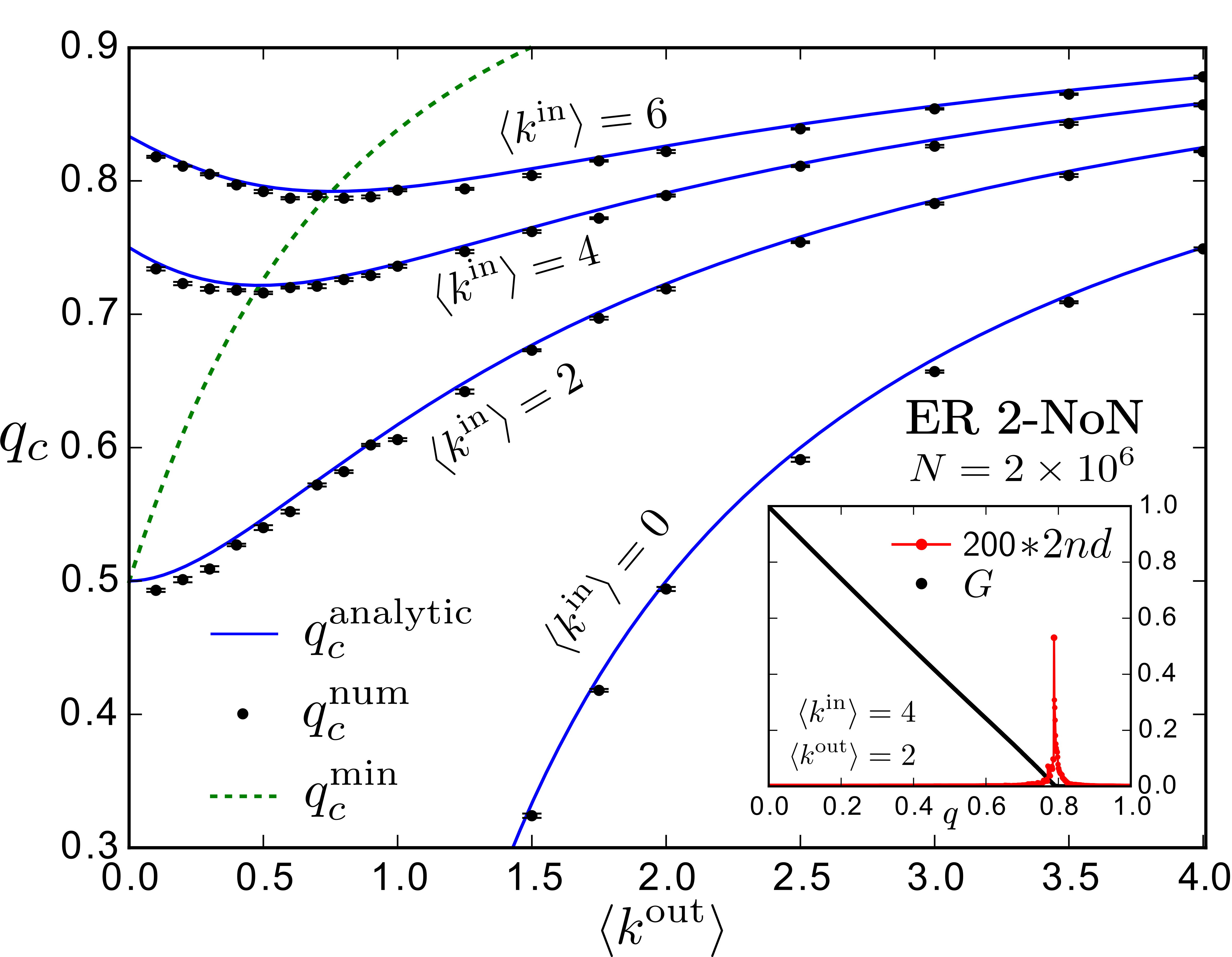}
\caption{(Color online) \textbf{Percolation phase diagram for ER 2-NoN.} 
Blue curves show our analytical prediction of the percolation
threshold, $q_c^{\rm analytic}$, as a function of $\la k^{\rm out}
\ra$ for different values of $\la k^{\rm in}\ra = 0,2,4,6$, obtained
from Eq.\,(\ref{eq:stabilityCondition}). Black dots show the measured
numerical percolation threshold, $q_c^{\rm num}$, from direct
simulation of the random percolation process, obtained at the peak of
the second largest connected activated component. 
{\color{black}The green dashed line indicates the maximal vulnerability $q_c^{\rm min}$}.
The percolation transition $q_c$ denotes the critical fraction of randomly removed
nodes at which $G(q_c)=0$
collapses. Errors are s.e.m.\ over 10 NoN realizations of system size
$N = 2\times10^6$. 
{\color{black} {\bf Inset.} 
Size of $G$ (black dots) and $200$\,$*$\,size of the second largest connected activated component (red dots)
as a function of $q$ for an ER 2-NoN with $\la k^{\rm in}\ra=4$, $\la k^{\rm out}\ra=2$ and 
$N = 2\times10^6$. The peak is at $q_c^{\rm num}=0.788$.}
\label{fig:percolationThreshold}}
\end{figure}

Results for $q_c(z, w) = 1-p_c(z, w)$ in \mbox{ER 2-NoN} are shown in
Fig.\,\ref{fig:percolationThreshold} and confirm the excellent
agreement between direct simulations of the random percolation process
on synthetic NoN and the theoretical percolation threshold calculated
from Eq.\,(\ref{eq:stabilityCondition}).\ The numerically measured
percolation thresholds, $q_c^{\rm num}(z, w)$, were obtained at the
peak of the second largest activated component {\color{black}(Fig.\,\ref{fig:percolationThreshold}\,\,Inset)}, measured relative to
the fraction of randomly removed nodes in synthetic \mbox{ER 2-NoN}. 
The analytical prediction of the percolation threshold, $q_c^{\rm
  analytic}(z, w)$, was obtained from the numerical solution of
Eq.\,(\ref{eq:stabilityCondition}).

The large values of $q_c$ in the percolation phase diagram confirm
that the NoN is very robust with respect to random node failures.  The
results indicate, for instance, that a fraction of more than 70\% of
randomly chosen nodes in an ER \mbox{2-NoN} with $\la k^{\rm in}\ra =
4$ can be damaged without destroying the giant connected activated
component $G$.  Moreover, the percolation transition, separating the
phases $G > 0$ and $G = 0$, is of second order in the robust NoN {\color{black}(Fig.\,\ref{fig:percolationThreshold}\,\,Inset)}.

Interestingly, the phase diagram reveals that, for a given average
in-degree $z$, the NoN exhibits maximal vulnerability $ q_c^{\rm
  min}(w^{*}, z) = 1 - p_c^{\rm max}(w^{*}, z)$ at a characteristic
average out-degree $w^{*}(z)$, indicated by the dip in the percolation
threshold $q_c$ in Fig.\,\ref{fig:percolationThreshold}.  
{\color{black}The equation determining $w^*(z)$ can straightforwardly be obtained via implicit differentiation of \mbox{$\lambda_{+}(p_c, z, w) = 1$}, using $\partial p_c/\partial w\,|_{w^*}=0$, where $p_c(z,w)$ is given by the solution of Eq.\,(\ref{eq:stabilityCondition}). The corresponding curve for $q_c^{\rm min}(w^{*}, z)$ is shown in Fig.\,\ref{fig:percolationThreshold}.}
Conceptually,
the dip in $q_c$ occurs as a consequence of the competition between
dependency and redundancy effects in the NoN.  Starting from vanishing
inter-module connections, the critical fraction $q_c$, and therefore
the robustness of the NoN, initially decreases slightly as the number
of dependency links in the NoN is increased. However, upon further
increasing the density of inter-module dependencies, the resilience of
the NoN increases again with increasing redundancy among the
dependency connections.

%-----------------------------------------------%
%                     Figure  2                      %
%-----------------------------------------------%

The underlying mechanism responsible for the robustness of the NoN is
best understood from the behaviour of the model in the limit $\la
k^{\rm in}\ra \rightarrow 0$, which corresponds to a bipartite network
equipped with our activation rule for $\s_i$, given by
Eq.\,(\ref{eq:sigma}).  The corresponding message passing equations,
$\varphi_{i\to j} = \s_i\big[1-\prod_{k\in\mathcal{F}(i)\setminus
    j}(1-\varphi_{k\to i})\big]$, are straightforwardly obtainable
from Eqs.\,(\ref{eq:messagePassingRho})\&(\ref{eq:messagePassingVarphi}), and can be seen to coincide with
the usual single network message passing equations by observing that
the activation state $\s_i$ can actually be replaced with the
occupation variable $n_i$ in this case (the reason is the following:
assuming node $i$ is present $(n_i = 1)$, $\s_i = 0$ implies that none
of $i$'s out-neighbors is present and so none of the incoming
inter-module messages can be non-zero either).
This property can of course directly be
obtained also from Eq.\,(\ref{eq:leadingEigenvalue}), 
which in the limit $z = 0$ implies
\begin{equation}
\lambda_{\pm}^{z=0} = \frac{p}{2}\,\big\{\, w\,\pm\, \sqrt{\, w^2\,}
\,\big\}\ \hspace{.2cm}\rightarrow\,\,\ p_c^{z=0} = 1/w\ .
\end{equation}
Therefore, the functioning of dependency links is well-defined even if
they connect nodes that do not belong to the giant
connected activated component within each module. In the model of
\mbox{Refs\,\cite{Buldyrev2010:Cascade, Gao2012:Interdependent}}, on
the other hand, inter-module links only exist if they connect nodes
that belong to the largest connected activated component in their own module. Hence,
it is impossible to construct the NoN from below $p_c$ (or above
$q_c$) using dependency links. In the present robust model, we can
construct the links even if the nodes are not in $G$, allowing us to
build the NoN from below $p_c$ using dependency connections. Thus, the
transition is well-defined from above and below the percolation
threshold.

%------------------------------------------------------------------------%
%                               CONCLUSION                                %
%------------------------------------------------------------------------%
%{\bf Conclusion.---}
In conclusion, we have seen that the robustness in NoN can be
understood to emerge if dependency links do not need to be part of the
giant connected activated component $G$ for their proper functioning.
In contrast to previously existing models of interdependent networks
\cite{Buldyrev2010:Cascade, Gao2012:Interdependent}, dependencies in
the robust NoN do not lead to cascades of failures.  The key point in
our model is that a node can be activated even if it does not belong
to $G$.
An example of the structure of NoN where the model applies is that of
the brain \cite{gallos, saulo, preprint,sigman}.
While in Ref.\,\cite{saulo} we
have shown that the model of \cite{Buldyrev2010:Cascade} becomes
robust when correlations in the dependencies are considered, here we
show that a local activation rule Eq.\,(\ref{eq:sigma}) akin to brain
control between modules defines a novel model of NoN which is robust
even without correlations. 
The effect of degree correlations on the robustness of the
NoN is to be investigated \cite{saulo}.
The model is straightforwardly
generalizable also to directed links and to dependency connections not
restricted to be only across modules, but also inside each module.

{\bf Acknowledgment.}
We acknowledge funding from NSF PHY-1305476, NIH-NIGMS
1R21GM107641, NSF-IIS 1515022 and 
Army Research Laboratory Cooperative Agreement Number W911NF-09-2-0053
(the ARL Network Science CTA).

\newpage
%------------------------------------------------------------------------%
%                      B I B L I O G R A P H Y                           %
%------------------------------------------------------------------------%

\end{document}